\documentclass{aa}
\usepackage{graphicx,natbib}
\bibpunct{(}{)}{;}{a}{}{,}
\usepackage{txfonts}
\def\msol{M_\odot}

\begin{document}

\title{The effect of a planet on the dust distribution in a 3D protoplanetary disk}
\author{L. Fouchet\inst{1,2}  \and S. T. Maddison\inst{3}
\and J.-F. Gonzalez\inst{1} \and J. R. Murray\inst{3}}

\institute{Universit\'e de Lyon, Lyon, F-69003, France,
Universit\'e Lyon 1, Villeurbanne, F-69622, France,
CNRS, UMR 5574, Centre de Recherche Astrophysique de Lyon,
\'Ecole Normale Sup\'erieure de Lyon, 46 all\'ee d'Italie,
F-69364 Lyon cedex 07, France\\
e-mail: {\tt fouchet@phys.ethz.ch, Jean-Francois.Gonzalez@ens-lyon.fr}
\and
ETH Hoenggerberg Campus, Physics Department, HPF G4.2, CH-8093 Zurich,
Switzerland
\and
Centre for Astrophysics and Supercomputing, Swinburne University
of Technology, PO Box 218, Hawthorn, VIC 3122, Australia\\
e-mail: {\tt smaddison@swin.edu.au, jmurray@astro.swin.edu.au}}
\offprints{L. Fouchet}

\date{Received ?? ??? ???? / Accepted ?? ??? ????}

\abstract{
{\it Aims:} We investigate the behaviour of dust in protoplanetary disks 
under the action of
gas drag in the presence of a planet. 
Our goal is twofold: to determine the spatial distribution of dust depending on grain size and planet mass, and therefore to provide a framework for interpretation of coming observations and future studies of planetesimal growth.
\\
{\it Method:} We numerically model the evolution of dust in a protoplanetary
disk using a two-fluid (gas $+$ dust) Smoothed Particle Hydrodynamics (SPH) 
code, which is non-self-gravitating and locally isothermal.  The code follows 
the three dimensional distribution of dust in a protoplanetary disk as it 
interacts with the gas via aerodynamic drag.  In this work, we present the 
evolution of a minimum mass solar nebula (MMSN) disk comprising 1\% dust 
by mass in the presence of an embedded planet. 
We run a series of simulations which vary the grain size and planetary mass 
to see how they affect the resulting disk structure.\\
{\it Results:} 
We find that gap formation is much more rapid and striking in
the dust layer than in the gaseous disk and that a system with a given
stellar, disk and planetary mass will have a completely different appearance
depending on the grain size. 
For low mass planets in our MMSN disk, a gap can open in the dust disk while not in
the gas disk.  We also note that dust accumulates at the external edge of the
planetary gap and speculate that the presence of a planet in the disk may 
enhance the formation of a second planet by facilitating the growth of 
planetesimals in this high density region.

}

\maketitle

\section{Introduction}

The effect of a planet in a gaseous disk has been well studied both
analytically and numerically  
\citep{GoldTremaine1979,GoldTremaine1980,PapLin1984,LinPap1986,LinPap1993,Kley1999,Bryden-etal1999,dValBorro-etal2006}.
Planetary gaps result from tidal interactions 
between a sufficiently massive planet and the disk if the tidal torques 
induced by the planet exceed the viscous torques of the disk.
If the gap is sustained, its presence severely decreases accretion of disk 
material by the planet  \citep{ArtyLub96,Kley1999,Bryden-etal1999}, 
which suggests that the planet formation process is self-limiting.  
\citet{Lufkin-etal2004} suggest, however, that the accretion rate is not so 
dramatically reduced in self-gravitating disks.
The exchange of angular momentum that results from the tidal torques leads to
planetary migration, which is slowed once a gap has formed.  This is known as 
Type~II migration and proceeds on a viscous timescale 
\citep{PapLin1984,Takeuchi-etal1996,Ward1997}.
If the planet mass is not massive enough to sustain a gap, it will undergo the faster Type~I 
migration \citep{GoldTremaine1979,Ward1997}, while intermediate-mass planets only open a partial gap and suffer Type~III, or runaway, migration \citep{MassetPap2003,Arty2004}.

There are a variety of observational signature of planetary gaps, including
infrared dips and direct scattered light and sub-millimetre observations of
protoplanetary disks. 
Mid-infrared dips are seen in quite a few pre-main sequence disks around 
T~Tauri stars including T~Tau and GM~Aur.
The mid-IR emission comes from the inner $1-10$~AU of 
the disk and the 10~$\mu$m feature is predominantly from silicate grain 
emission and it has been suggested that such IR dips may be the results 
of an unseen planet clearing out the disk material 
\citep[e.g.][]{Calvet-etal2002,Rice-etal2003,Varniere-etal2006a}.
Such dips, however, have also been modelled assuming a continuous disk with 
a more realistic density distribution than a standard power-law 
\citep{BossYorke1996}, and by a continuous disk with silicate grain growth 
\citep{DullDominik2005}.
Sub-millimetre observations of older debris disks show brightness asymmetries
and a range of structures within the disk, many of which can be modelled by 
the presence of an unseen planet locking the dust in resonances  
\citep{Ozernoy-etal2000,Wilner-etal2002,DellerMaddison2005}. However 
it has been noted by \citet{Wyatt2006} and \citet{Grigorieva-etal2006}
that holes and clumpiness of debris disks can result purely from collisional cascade of 
planetesimals within the disk. 
\citet{TakeuchiArty2001} show that gas drag alone can be used to explain annulus-like disks with sharp edges, and
sub-mm holes have also been explained by the 
sublimation of icy grains  \citep{Jura-etal1998}.
Direct imaging of scattered light from protostellar and debris disks with
HST confirm the presence of gaps 
\citep{Weinberger-etal1999,Schneider-etal1999,Ardila-etal2004,Kalas-etal2005,Schneider-etal2006}

Recent models have indicated that ALMA will be able to detect planetary gaps
at sub-millimetre wavelengths. Using the results of 2D hydrodynamics 
simulations, \citet{Wolf05} use a Monte Carlo radiative transfer code to 
predict the emission of a disk hosting a planetary gap carved out by a range 
of planet masses.  They find that ALMA will be able to detect the gap in a 
$0.01 M_{\odot}$ disk with an $M_{\rm Jup}$ planet at a distance of 100~pc. 
When using 2D hydrodynamics codes, one needs to add an analytic disk 
atmosphere in order to provide the Monte Carlo simulation with a 3D density 
structure. \citet{Wolf05} assume an un-flared disk, while \citet{Varniere-etal2006a}
use a flared disk and show that the spectral energy distribution and
synthetic scattered light images are effected by the outer wall of the gap.
Both groups, however, assume that the gas and dust are well-mixed within the 
disk and yet the gas-to-dust ratio will change substantially in protoplanetary
disks as the grains begin to grow and settle to the midplane 
\citep{Maddison-etal2003}.

In recent years, a growing number of studies started to consider the separate
evolution of gas and dust in disks. \citet[][hereafter BF05]{BF05} 
described the simultaneous radial migration and vertical settling of dust
grains, treated as a separate fluid in global 3D SPH simulations, and their
dependency on grain size. \citet{Rice-etal2004} showed that large solid bodies,
treated as trace particles, tend to accumulate in the spiral arms in massive
disks. The influence of the magneto-rotational instability 
\citep[MRI,][]{BalbusHawley1991} has also been studied in Eulerian global 2D 
simulations \citep{FromangNelson2005} and local 3D simulations 
\citep{FromangPap2006,JohansenKlahr2005,Johansen-etal2006}. A fraction of the 
solid bodies, treated as a separate fluid for small sizes or as particles for 
large sizes, were found to be trapped in density maxima generated by the MHD turbulence.

As well as the gas-to-dust ratio varying throughout the disk, we expect that
the effects of planetary gaps will be stronger in the dust phase than in the 
gas phase, which will further affect observations. 
Any gap created by a planet will result in a pressure minimum in the disk, and
unless the dust and gas are extremely well coupled and the motion is governed 
by the gas (which will be true for sub-micron sized grains), the gas will
respond to the pressure gradient while the dust will not.
(The dust feels no viscous forces and much lower pressure forces than
the gas and hence the tidal torques which create the gap will dominate.)  
If anything, the dust moves to regions of pressure maxima
\citep[see][]{Haghi03} on the edge of the gap. Furthermore, it is easier to
form gaps in 2D disks than in 3D disks \citep[see][]{Kley1999} 
and we expect the dust to settle to the midplane on a much shorter timescale 
than the gas phase. Both of these phenomena can occur at the same time.
Thus predictions of observations of gap which assume a well-mixed
disk are probably too pessimistic and it is therefore important to study the 
effects of a planet on the dust layer to help constrain the characteristics 
of the planets we might detect via the gaps they create.

While many semi-analytic and numerical studies have investigated the formation
of planetary gaps in gas disks, very little work had been done on the effect of a 
planet in a dusty disk. 
Over the last few years, however, attention has begun to focus of the dust
phase of 2D protoplanetary disks with embedded planets. 
\citet{Paardekooper04,Paardekooper06}
studied the conditions for gap formation and showed that a smaller
planetary mass is required to open a gap in the dust disk rather than in the
gas. They note, however, that one should be careful when interpreting results 
for planets lighter than $0.1 M_{\rm Jup}$ in two dimensional simulations.
\citet{Rice-etal2006} found that the pressure gradient at the outer gap edge
can act as a filter keeping large grains out of the inner disk, which may
explain the deficit of near-IR flux observed in the SED of some T Tauri stars.
\citet{AlexanderArmitage2007} studied the dynamics of dust and gas during disk
clearing and, adding an embedded planet, show that observations of transition
disks can help discriminate between different models of disk clearing.

In this paper we study the formation of a gap triggered by an embedded planet 
in the dust layer of a protoplanetary disk. We will study the effects of planet
mass and grain size on gap formation and evolution in 3D, two-phase 
(gas$+$dust) protoplanetary disks.
These result will be compared with the gas-only simulations of others 
who must then assume the gas and dust are well mixed in order to determine how the
dust-gas coupling (or decoupling) affects potential observations of planetary
gaps.


\section{Gap opening criteria}
\label{sec-gapcriteria}

The gravitational perturbation of the planet launches a spiral density wave at 
the Lindblad resonances of the disk. The resulting tidal torques lead to an 
exchange in angular momentum when the waves damp by viscous dissipation or 
shocks.
Waves exterior (interior) to the planet carry excess (deficit) angular 
momentum compared to the disk and hence when the wave damps, the disk gains 
(loses) angular momentum and moves outwards (inwards).  This results in the 
formation of a gap around the planet.  

In order to sustain the gap, there needs to be a balance
between the tidal torques, which act to clear the gap, and the viscous
torques, which act to fill the gap \citep{LinPap1979}.
The gap criterion \citep{PapLin1984,Bryden-etal1999} is given by 
\begin{eqnarray}
\tau_{\rm tidal} & > & \tau_{\rm visc} \nonumber \\
\frac{M_{\rm p}}{M_{\star}} & > & 40 \alpha_{\rm SS}  
                \left(\frac{H}{r_{\rm p}}\right)^2 \, ,
\label{eqn-gapvisc}
\end{eqnarray}
where $M_{\rm p}$ and $M_{\star}$ are the mass of the planet and star, 
$\alpha_{\rm SS}$ is the \citet{ShakuraSunyaev1973} viscosity parameter, $H$ 
is the disk scale height and $r_{\rm p}$ is the semi-major axis of the planet.
For inviscid disks, the competing forces are local pressure and gravitational 
tidal perturbations \citep{LinPap1993} and gap criterion is given by
\begin{eqnarray}
F_{\rm tidal} & > & F_{\rm press}  \nonumber \\
 \frac{M_{\rm p}}{M_{\star}}  & > &  3 \left(\frac{H}{r_{\rm p}}\right)^3 \, .
\label{eqn-gapinvisc}
\end{eqnarray}

Our previous simulations  (BF05)  show that the thickness of the dust layer 
depends on the grain size because different sized grains fall to the midplane 
at different rates.  
The gas drag ensures that (for certain grain sizes which 
depend on the nebula conditions) the dust rapidly settles to the midplane 
and since the above expressions for the gap criteria depend on the disk 
scale height, this would suggest that it is easier to create and sustain a 
gap in the dust layer than in the gas.
However, Eqs.~(\ref{eqn-gapvisc}) and (\ref{eqn-gapinvisc}) were derived without taking the drag force into account.  Very small grains, which feel the strongest drag, are basically co-moving with the gas and are not likely to create a much sharper gap than the gas. Conversely very large grains (or boulders) are essentially decoupled from the gas and, feeling very little drag, have very long settling times. In this paper, we concentrate on the more interesting intermediate sized grains with the most efficient settling rates, for which the situation is greatly complicated by the drag forces and therefore  numerical simulations are necessary to fully describe their behaviour.
The regime we consider here is somewhat different to the 2D work of  \citet{Paardekooper04} who study dusty disks in which the dust is strongly coupled to the gas and responds indirectly to the planet gravity through small radial pressure gradients in the gas disk that leads to the formation of a gap.


\section{Code description}

We have developed a 3D, two-fluid (gas$+$dust), locally isothermal, 
non-self-gravitating code based on the Smoothed Particles Hydrodynamics (SPH) 
algorithm, a Lagrangian technique described by \citet{Monaghan92}. 
The dusty-gas is approximated by two inter-penetrating
fluids that interact via aerodynamic drag.  We assume Epstein
drag, which is valid for the range of dust sizes and nebula parameters used in 
this study.  Thus the drag force $F_\mathrm{D}$ is given by
\begin{equation}
F_\mathrm{D}=\frac{4 \pi}{3} \rho_\mathrm{g} s^2 v c  \, ,
\end{equation}
where  $\rho_\mathrm{g}$ is the gas density, $s$ is the (spherical) grain
radius, $c$ is the sound speed, and
$v=|\vec{v}_\mathrm{d}-\vec{v}_\mathrm{g}|$ is the velocity difference between
dust and gas, given by $\vec{v}_\mathrm{d}$ and $\vec{v}_\mathrm{g}$
respectively 
\citep{Stepinski96}.
The dust particles are incompressible and there is no
coagulation or evaporation of grains. All simulations in this paper are 
conducted with just one grain size at a time. In future work we shall 
include multiple grains sizes per simulation.
For details of how the equations of motion and the density of
the two fluids are calculated, we refer the reader to BF05.

\subsection{Implementation and initial conditions}

We set up a disk of gas and dust with a total mass of $M_{\rm disk}$ around a 
one solar mass star into which we embed a planet of mass $M_{\rm p}$ at a
distance of $r_{\rm p}$.  The system is evolved for about 100 planetary 
orbits.

The code units are chosen such that $G=M_{\star}=r_{\rm p} = 1$.
The disk equation of state is isothermal and 
hence the temperature is constant in the vertical direction but
follows a radial temperature profile 
($T \propto r^{-1}$).
The isothermal
sound speed at $r=1$ is $c = H/r = 0.05$, where $H$ is the disk scale height 
and $r$ the radial distance from the star.  Any heat produced in the disk due 
to tidal forces or viscous dissipation is assumed to be immediately radiated 
away.  
In order to compare our results to those of  \citet{Paardekooper06}, we use an
initially constant surface density profile. In the $z$ direction particles are
randomly distributed between $\pm H$. 
The intrinsic density of the dust grains, $\rho_d$, is  
1.25~g~cm$^{-3}$.

All simulations contain an equal number of gas and dust particles and the
disks contain 1\% dust by mass. We start our simulations with a gas disk
containing 50,000 particles and an embedded planet. This is evolved for 8
planetary orbits before we add the dust phase.  The dust particles are
initially overlaid on the gas particles and the system is then evolved for a 
further 96 orbits, making 104 orbits in total.  
Given that some accretion occurs during the first 8 orbits, we generally end up 
with about 96,000 particles in total. While this may seem to be quite low resolution 
for today's SPH simulations (c.f. $N_{\rm gas}\sim$300,000 in \citeauthor{Rice-etal2003}
\citeyear{Rice-etal2003} and \citeauthor{Schafer-etal2004} \citeyear{Schafer-etal2004})
BF05 showed that this level of resolution gives the same results as simulations 
with almost twice as many particles (in the case of a disk without a planet). It
should be noted that in our two-phase code, the time step is severely limited
by the drag timescale. While the drag term is solved implicitly to alleviate
the cripplingly short $dt_{\rm drag}$ issue, this requires an extra $N/2$
operations per time step (and an additional SPH particle neighbour search) 
and thus particle resolution suffers as a consequence. 
To ensure the validity of our 
low resolution results, we ran a series of high resolution simulations with 
$N$$\sim$400,000 which are presented in Sect.~\ref{subsec-standardmodel}.

The planet is implemented as a point mass particle which moves under the
gravitational influence of the central star, but does not feel pressure,
viscous or drag forces.  The SPH particles, on the other hand, feel the
gravitational potential of the planet.  The planetary orbit can either be
circular, elliptic or even hyperbolic and is determined by the choice of the
initial velocity. In this work we only study circular orbits.  Particles are
accreted by the planet if they cross a critical radius given by  
\begin{equation}
R_{\rm crit} = 0.1R_{\rm Hill}=r_{\rm p} \left( \frac{M_{\rm p}}{3M_{\star}}  \right)^{1/3}.
\end{equation}
We take no account of any mass or angular momentum transfer corresponding 
to this infall and therefore do not simulate migration of the planet.
The central star also accretes particles that get closer than $r_{\rm in}=0.4$ code units, 
which sets the inner edge of the disk. This rather large value was chosen to 
compare with the results of \citet{Paardekooper04,Paardekooper06}. 
(See,
however, Sect.~\ref{subsec-boundary} for a discussion about the location 
of the inner boundary and its effect on the gap.)
Particles are also removed from the simulation if they escape to large radii, 
which is set to $r_{\rm esc}=4$ code units.

\subsection{Simulation Suite}

In this work we present the results of a small, low mass disk close to 
the minimum mass solar nebula (MMSN).  The disk mass is set to 
2.9 10$^{-3}$~$\msol$.
Given the code scaling, $R_{\rm Hill} =  0.36$~AU, the stellar accretion radius is $r_{\rm in} = 2$~AU,  
and $r_{\rm esc} = 20.8$~AU, with the latter two setting the radial extent of the disk.
Our standard model has a 1 Jupiter mass ($M_{\rm Jup}$) planet on a circular 
orbit of radius 5.2~AU in a disk containing grains 1~m in size.  
Such a large grain size was chosen for our standard model because for these nebula conditions this is the grain size that settles most efficiently and thus the effect of the planet is most striking in a dust layer of 1~m sized ``boulders''. Further, boulders of this size should radially migrate  most rapidly and here we will show that this migration can 
be stopped by planetary gaps.  While the 1~m grains are observationally uninteresting, a large population of boulders would result in a population of smaller grains by collisions which are detectable. 

We start by running our standard MMSN model and compare the evolution 
of the planetary gap in
the gas and dust phases. We then run a series of experiments to study the 
effect of grain size in the dust disk, with $s =$ 1~cm, 10~cm 
and 1~m.
This is followed by a series of experiments that study the effects of planetary 
mass on gap formation and evolution, with $M_{\rm p}=0.05, 0.1, 0.2, 0.5$ and 
1.0 $M_{\rm Jup}$.
The parameters of our simulation suite are summarised in Table~\ref{table-1}. 

\begin{table}
\caption{Parameters of our simulation suite.}
\label{table-1}
\centering
 \begin{tabular}{llll}
 \hline\hline
\multicolumn{4}{c}{MMSN models} \\ \hline
 Star mass & 1 $\msol$ & Inner radius & 2~AU  \\
 Disk mass & 2.9 10$^{-3}$~$\msol$ & Outer radius & 20.8~AU \\
\hline
$s$ (m) & $M_p$ ($M_{\rm Jup}$)  & $\alpha$, $\beta$ & Drag \\
\hline
 1.0  & 1.0  & 0.1, 0 & yes \\
\hline
 0.1  & 1.0  & 0.1, 0 & yes \\
 0.01  & 1.0  & 0.1, 0 & yes \\
\hline
  1.0  & 0.5  & 0.1, 0 & yes \\
 1.0  & 0.2  & 0.1, 0 & yes \\
 1.0  & 0.1  & 0.1, 0 & yes \\
 1.0  & 0.05   & 0.1, 0 & yes \\
\hline
 1.0  & 1.0  & 0.1, 0.5 & yes \\
 1.0  & 1.0  & 1, 2 & yes \\
\hline
 1.0  & 1.0  & 0.1, 0 & no \\
 1.0  & 1.0  & 0.1, 0 & yes \\
\hline\hline
\end{tabular}
\end{table}

\subsection{Dust stopping time}

Since the drag force acting on the grains depends on the nebula gas density and relative velocity 
as well as the grain size, it is useful to know the stopping time for different grain sizes in order to 
compare with different nebula models.
The stopping time \citep{Stepinski96} in the Epstein drag regime (assuming solid spherical grains) is given by 
\begin{equation}
t_{\rm s} = \frac{m v}{F_{\rm D}} = \frac{\rho_{\rm d}  s} {\rho_{\rm g}   c}  
\end{equation}
where $\rho_{\rm d}$ is the intrinsic dust density.
This is made non-dimentional via $t_{\rm s} = T_{\rm s} / \Omega_{\rm Kep}$, and thus the 
non-dimensional stopping time, $T_{\rm s}$, can be written as
 \begin{equation}
          T_{\rm s} =  \frac{\rho_{\rm d}  s  \Omega_{\rm Kep}} {  \rho_{\rm g}   c} \, .
\label{eqn-stoppingtime}          
\end{equation}
where $\Omega_{\rm Kep}=v_{\rm Kep}/r$ is the Keplerian angular velocity.
We find that at the location of the planet for our standard model  (see Sect.~\ref{subsec-standardmodel}) 
the stopping time is given by 
$T_{\rm s} =  0.067 s_{\rm  cm}$ where $s_{\rm cm}$ is the grain size in cm.  
Thus $T_{\rm s} = 0.067, 0.67$ and 6.7 for 1~cm, 10~cm and 1~m grains respectively.  
This is in good agreement with \citet{Paardekooper06} (noting that they have an additional factor of $\sqrt{\pi/8}$ in their expression of the stopping time, following \citeauthor{TakeuchiLin2002} \citeyear{TakeuchiLin2002}).



\begin{table}
\caption{Percentage of particles lost due to accretion or ejection for various SPH 
viscosity parameters $\alpha$ and $\beta$. }
\label{table-2}
\centering
\begin{tabular}{ccc}
\hline\hline
\multicolumn{2}{c}{viscosity parameters} & {\% of particles } \\
$\alpha$ & $\beta$ & lost \\
\hline
0.1 & 0   & 16.1~\% \\
0.1 & 0.5 & 16.6~\% \\
1   & 2   & 32.5~\% \\
\hline
\end{tabular}
\end{table}

\begin{figure}
\includegraphics[width=8cm]{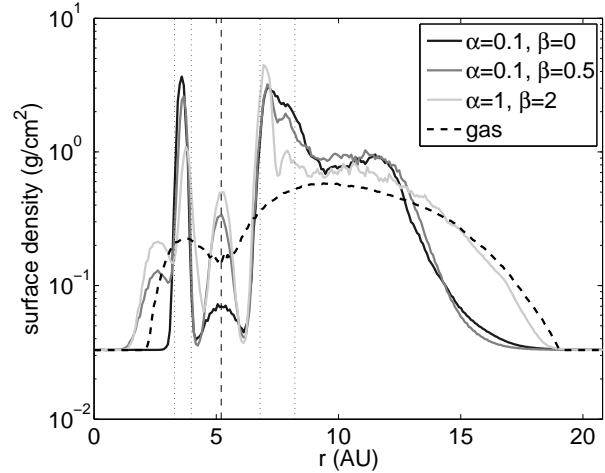}
\caption{Comparison of the dust disk surface density after 104 planet orbits
  for a range of SPH viscosity parameters, $\alpha$ and $\beta$, for 
  our standard model with a 1~M$_{\rm Jup}$ planet at 
  5.2~AU.  The gas surface density is scaled by a
  factor of 0.01 for direct comparison with the dust phase. The vertical lines 
  are, from left to right,
  the 1$:$2 and 2$:$3 internal resonances, the 1:1 planetary orbit, and the
  3$:$2 and 2$:$1 external resonances respectively.}
\label{fig-viscosity}
\end{figure}

\subsection{Viscosity}

Since protoplanetary disks are viscous accretion
disks, we can apply the Shakura-Sunyaev $\alpha$-disk prescription 
\citep{ShakuraSunyaev1973}
in which the kinematic viscosity is given by 
\begin{equation}
\nu = \alpha_{\rm SS} c H \,.
\label{eqn-SSalpha}
\end{equation}
In the thin disk approximation, $H/r = c/ v_{\rm Kep}$ and the 
viscosity can be rewritten as $\nu = \alpha_{\rm SS} c^2 / 
\Omega$, and $\alpha_{\rm SS}$ is close to 0.01 
\citep[which provides a good fit to observations of
protostellar disks -- see ][]{Hartmann-etal1998,King-etal2007}. 

We use, as in BF05,
the standard SPH artificial viscosity term  \citep{JJMBG83,Monaghan89} 
that is included in the SPH momentum equation and acts as a viscous 
``pressure''. Using the SPH particle subscript notation $i$ and $j$, the 
viscous term of the momentum equation is given by
\begin{equation}
\frac{d {\bf v}_i}{dt} = - \sum_j m_j \Pi_{ij} \nabla_i W_{ij} \, ,
\label{eqn-SPHviscmomentum}
\end{equation}
where ${\bf v}$ is the particle velocity, $m$ the particle mass, $W$ the 
smoothing kernel, and $\Pi$ the artificial viscosity term.
The most common form of $\Pi_{ij}$ is due to \citet{Monaghan89} and is given 
by 
\begin{equation}
\Pi_{ij} =   -\frac{\alpha \overline{c}_{ij} \mu_{ij} + \beta \mu_{ij}^2}
             {\bar{\rho}_{ij}} \,  ,
\label{eqn-SPHviscosity}
\end{equation}
where 
$\mu_{ij} = {h{\bf v}_{ij} \cdot {\bf r}_{ij}}/ ({r_{ij}^2 + \eta^2})$
if ${\bf v}_{ij} \cdot {\bf r}_{ij} < 0$ and is zero otherwise, 
${\bar{\rho}_{ij}}$ and $\bar{c}_{ij}$ are the average density and average 
sound speed of interacting particles $i$ and $j$, and $\eta^2 = 0.01 h^2$ is a 
softening parameter that prevents $\mu_{ij}$ becoming singular, and $h$ is 
the SPH smoothing length.
We use the cubic spline kernel of Monaghan \& Lattanzio (1985) given by
\begin{eqnarray}
W({\bf r},h) = \frac{\sigma}{h^D}
     \left\{
        \begin{array}{lll}
        1-\frac{3}{2}(r/h)^2 + \frac{3}{4}(r/h)^3 \, , 
                        & \mbox{if $0 \le (r/h) \le 1$; }\\
        \frac{1}{4}(2-(r/h))^3 \, ,                
                        & \mbox{if $1 \le (r/h) \le 2$;  }\\
        0    \, ,                              
                        & \mbox{otherwise, }
        \end{array}
     \right.
\label{eqn-SPHkernel}
\end{eqnarray}
where $D$ is the number of dimensions, and ${\sigma}$ is a normalisation 
constant given by $2/3, 10/7\pi$ and $1/\pi$ in one, two and three dimensions 
respectively.

There are two free SPH viscosity parameters, $\alpha$ and 
$\beta$, which control the strength of the viscosity. (Note that 
the Shakura-Sunyaev $\alpha_{\rm SS}$ and the SPH $\alpha$ are not the same.) 
The viscosity associated with the linear $\alpha$ term of $\Pi_{ij}$ was 
introduced to remove subsonic velocity oscillations that follow shocks 
\citep{JJMBG83}, while the nonlinear $\beta$ term damps high 
Mach number shocks and prevents particle interpenetration \citep{Monaghan89}. 
While it is reasonable to set $\beta$ to zero since we do not expect 
strong shocks in our protoplanetary disk simulations \citep[c.f.][]{Bryden-etal1999}, 
some artificial bulk viscosity via the $\alpha$ term is needed to 
prevent random velocities from growing and producing a thick disk. 
Substituting Eqs.~(\ref{eqn-SPHviscosity}) and (\ref{eqn-SPHkernel}) into 
Eq.~(\ref {eqn-SPHviscmomentum}), it can be shown that $\alpha$ is 
related to the viscosity via
\begin{equation}
\nu_{\rm sph} = \tilde{\sigma} \alpha c h \, ,
\label{eqn-SPHalpha}
\end{equation}
where $\tilde{\sigma}=$ 1/8 for a 2D cubic spline kernel or 2/15 for a 3D 
cubic spline kernel \citep{Pongracic88,Murray96}.
Equating Eqs.~(\ref{eqn-SSalpha}) and (\ref{eqn-SPHalpha}), we can relate
$\alpha_{\rm SS}$ and $\alpha$ for our 3D simulations via
\begin{equation}
\alpha_{\rm SS} = \alpha \left( \frac{\tilde{\sigma} h \Omega}{c}\right) \, .
\end{equation}
In order to keep the corresponding Shakura-Sunyaev viscosity parameter
$\alpha_{\rm SS}$ close to 0.01 
\citep[as indicated by observations of protoplanetary disks 
- see][]{Hartmann-etal1998,King-etal2007}, 
we set the $\alpha = 0.1$.
A problem with using SPH to model  planetary gaps is that the numerical 
viscosity cannot be reduced to arbitrarily small values, which means that 
(gas) particles will diffuse into the gap.
Many numerical simulation of planetary gaps use grid-based codes
\citep[e.g.][]{DAngelo2002} in order to keep $\alpha_{\rm SS}$ as low as possible.
Some grid codes can get $\alpha_{\rm SS}$ to be as low as $10^{-4}$. 
However, since observations of protoplanetary disks indicate $\alpha_{\rm SS} \sim 0.01$, 
SPH can do an adequate job and indeed one of the first simulations of an
embedded planet in a disk by \citet{SMH87} used SPH.
\citet{Schafer-etal2004}
present a new treatment of the viscosity in SPH codes which includes an
additional artificial bulk viscosity term in the Navier-Stokes equation and 
demonstrate that their results compare well with grid-based codes. 
This is however very 
computationally expensive and has not been implemented in our code.

To check the validity of the SPH viscosity parameters, we ran a series of
tests to investigate the effect of changing $\alpha$ and $\beta$. Our standard
model has $\alpha=0.1$ and $\beta=0$, and we also ran test 
simulations with ($\alpha,\beta$)=(0.1,0.5) and (1,2). 
As expected, we find the ($\alpha,\beta$)=(1,2) case to be much
more dissipative than the two other.  To quantify this, we
calculate the percentage of particles lost between the 8th~orbit (when the
dust phase is incorporated) and 104th~orbit (the end of the simulation). 
Particles can be lost either by accretion onto the central star, accretion onto
the planet or escape to large radii.  The results of these tests are given in
Table~\ref{table-2}. We find that about twice as many particles are lost when
($\alpha,\beta$)=(1,2) than in the other cases. Simulations with
($\alpha,\beta$)=(0.1,0) and (0.1,0.5) give similar results, with the latter
being a little more dissipative. The evolution of the
disk with different SPH viscosity parameters is shown in Fig.~\ref{fig-viscosity}.
We can see that the ($\alpha,\beta$)=(1,2) simulations gives qualitatively different 
results, and that the ($\alpha,\beta$)=(0.1,0) and (0.1,0.5) simulations differ 
mainly in the corotation region and close to the edges of the gap where shock 
capture is more important. 

In order to keep the viscosity as low as possible, and considering that 
results with $\beta = 0$ are not so different from those with $\beta = 0.5$, we keep 
$\alpha=0.1$ and $\beta=0$ for the remainder of our simulations.

\section{Simulation results}

Here we present the results from our suite of simulations, starting with our 
standard disk model, followed by our grain size tests and finally planetary mass tests.

\subsection{Standard model}
\label{subsec-standardmodel}
In the standard model, we set up a 2.9 10$^{-3}$~$\msol$ disk containing 1\% dust 
by mass made up of boulders 1~m in size. The disk extends from 2~AU to 20~AU around 
a 1~$\msol$ star and hosts a $1 M_{\rm Jup}$ planet in a circular orbit at 5.2~AU. 

\begin{figure}[h]
\resizebox{\hsize}{!}{
\includegraphics{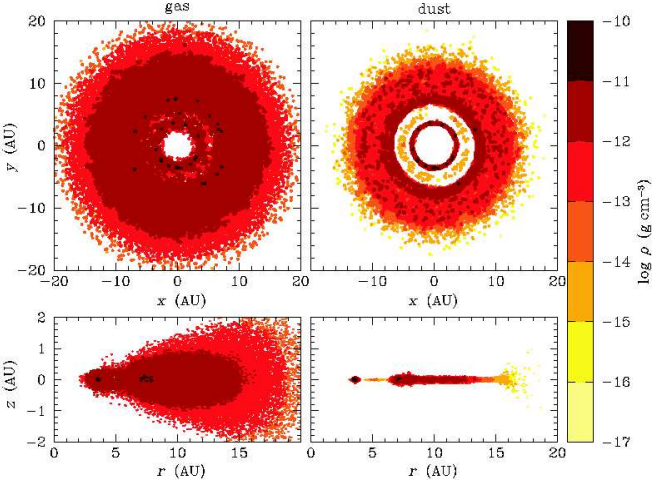}
}
\caption{End state of the standard model after 104 orbits (1,230~years). 
    Top: top-down view. Bottom: side-on view. Left: gas disk coloured by density. 
    Right: dust disk coloured by density.}
\label{fig-MMSN-standard3D}
\end{figure}

In Fig.~\ref{fig-MMSN-standard3D} we show the top-down ($x,y$) and side-on ($r,z$)
view of the density structure in the gas and dust phases of the standard 
model. 
While the planet opens a shallow gap in the gas disk, the gap it creates in the 
dust layer is much more striking. 

\begin{figure*}
{
\includegraphics[width=8cm]{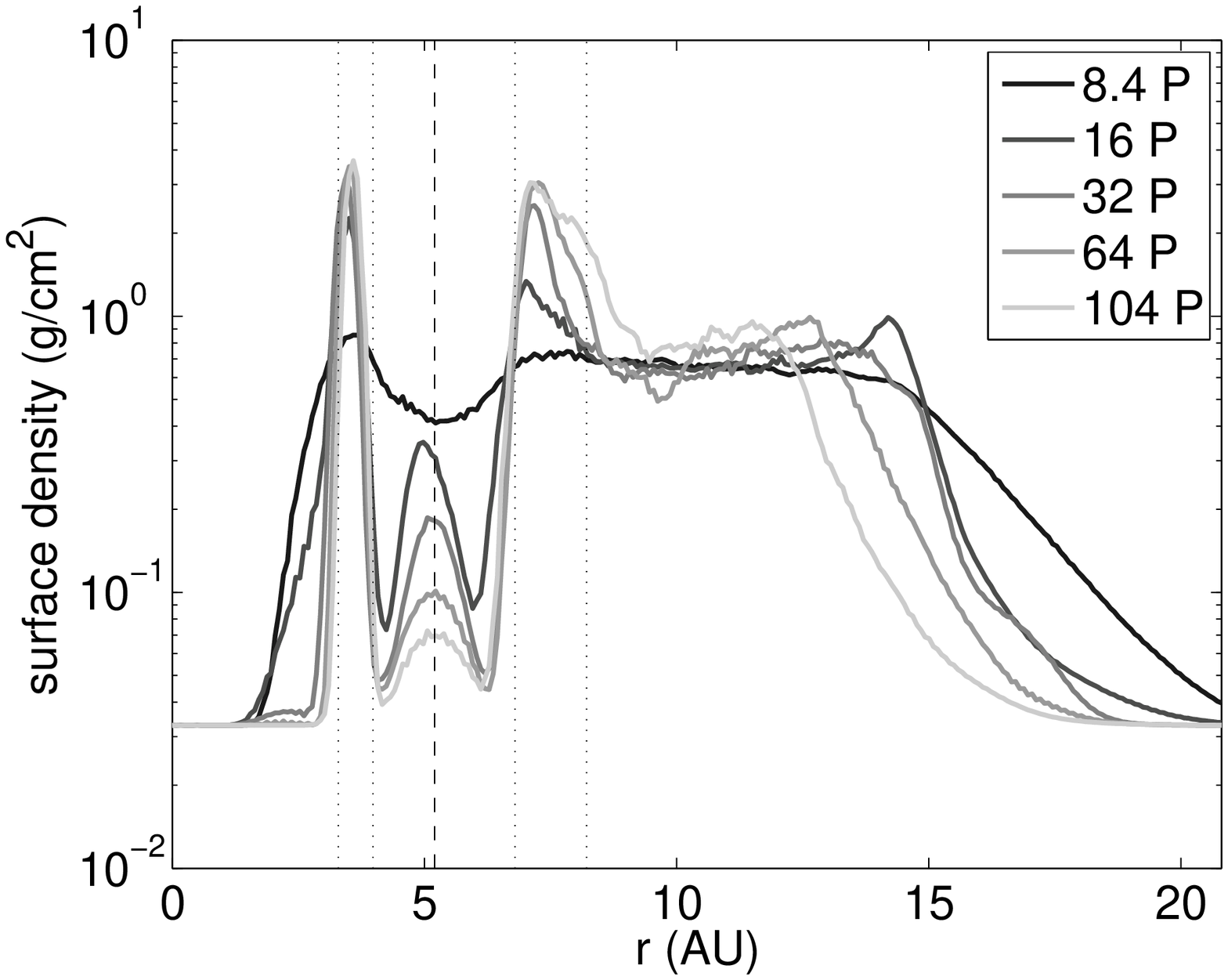}
\includegraphics[width=8cm]{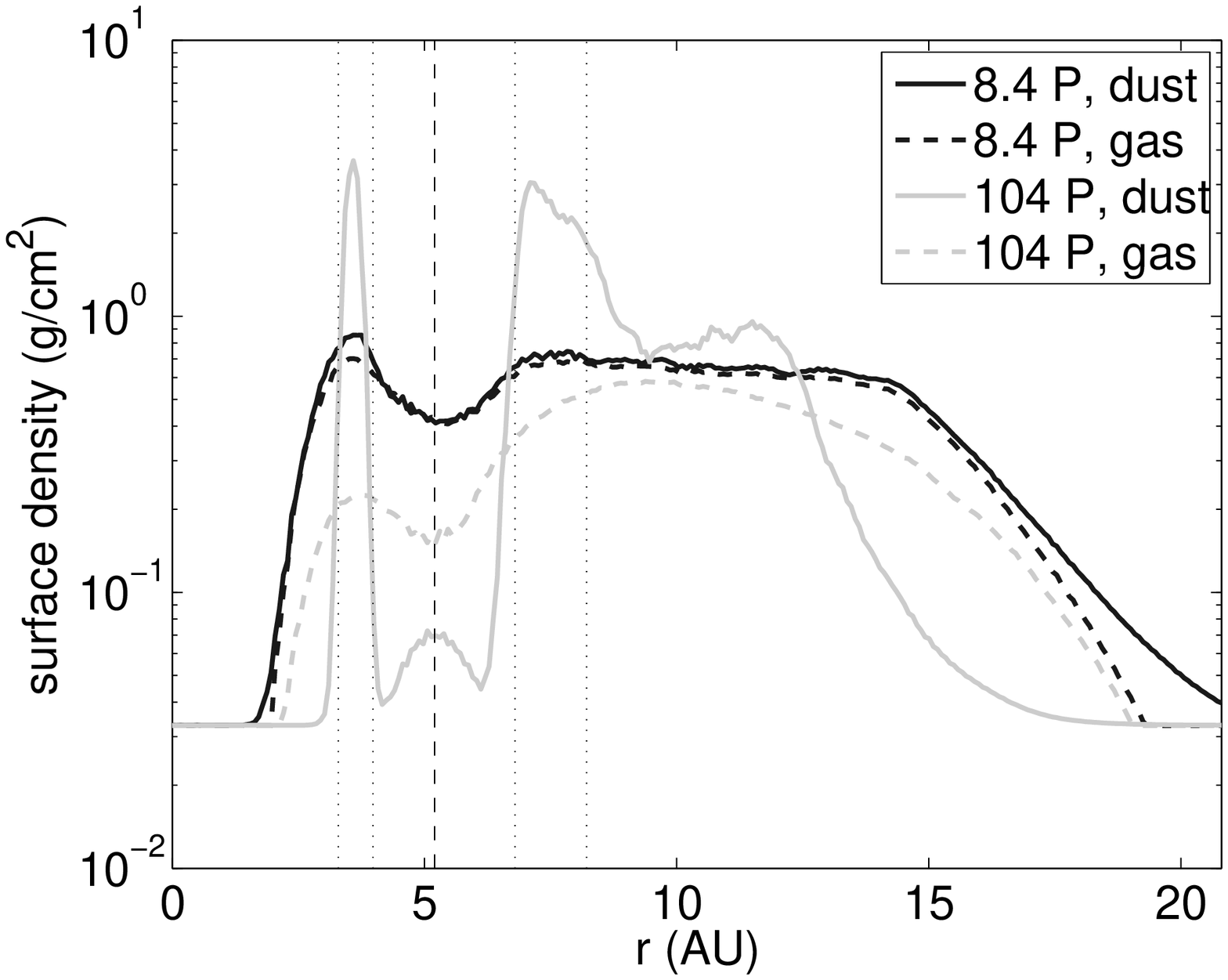}
}
\caption{Azimuthally averaged surface density profiles for a gap created by a
  1~$M_{\rm Jup}$ planet at 5.2~AU in the standard model. The left panel shows the
  evolution of the 1~m sized grain dust layer after 8.4, 16, 32
  and 64 and 104 planet orbits.  The right panel shows the comparison of the
  gas and dust surface densities after 8.4 and 104 orbits. The gas density is
  scaled by 0.01 for direct comparison to the dust density.
  The vertical lines are the resonances as described in Fig.~\ref{fig-viscosity}. }
\label{fig-MMSN-standard-2Ddens}
\end{figure*}

In Fig.~\ref{fig-MMSN-standard-2Ddens} we plot the radial profile of the azimuthally averaged surface
density of the standard model at five evolutionary stages ($8.4P,
16P, 32P, 64P$ and $104P$, where $P$ is one period of the planet).
Note that the dip in the radial density profile due to the planet induced gap
is surrounded by two density enhancements. The density enhancements are much 
sharper in the dust than in the gas even though the external enhancement gets 
broader over time due to feeding by external disk material. 
Particles that migrate inwards from the outer regions of the disk become trapped at the outer gap edge and over time more and more dust moves inwards and piles up at the outer gap edge, forcing this region of enhanced dust surface density to broaden.  
In the outer part of 
the disk, particles seem to be trapped by the 3$:$2 resonance as noted by
\cite{Paardekooper04,Paardekooper06}. We will return to this issue in 
Sect.~\ref{subsec-resonance}.
In the inner parts of the disk, on the other hand, particles flow through the
resonances without becoming trapped due to the combined effects of the spiral
waves created by the planet, the gas flow accreted by the central object, and the drag force
that causes the dust to drift inwards faster than the gas once it has collapsed to
the midplane. 
The internal over-density seen in Fig.~\ref{fig-MMSN-standard-2Ddens} 
cannot be explained by an accumulation of inwardly migrating grains, since they 
cannot cross the gap. It is instead due to the accumulation of dust at the gas pressure maximum.

The gap is not exactly centred on the planet's orbit but it is shifted towards the inner disk. This phenomenon was also seen in the simulations of \cite{Paardekooper04} and is due to the fact that the dust does not directly feel the shock related to the creation of the gap.  The shock only influences the dust via gas drag in the post-shock region. The flux of grains through the spiral wave is larger than if they were perfectly coupled to the gas and because the wave gradually depletes the gap, the density dip is deeper in the dust disk than in the gas. 

\begin{figure}[h]
\resizebox{\hsize}{!}{
\includegraphics{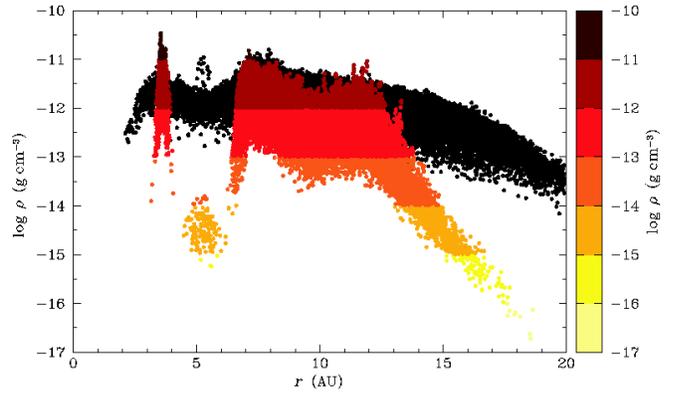}
}
\caption{Volume density profile of the gas and dust of the standard model after 104 orbits. 
    The black points represent the gas particles and the coloured points represent the dust particles 
    (coloured by density). Note that in this case the gas density has not been rescaled.}
\label{fig-MMSN-rho}
\end{figure}

Figure \ref{fig-MMSN-rho} shows the volume density of gas and dust as a function of distance to the star
at the end of the simulation. The general radial dependence of the volume density is
quite similar to that of the surface density seen in Fig.~\ref{fig-MMSN-standard-2Ddens}, with 
similarities seen in the shape of the dust profile around the gap and the accumulation of dust at 
the corotation radius.  Indeed, dust particles tend to pile up where the gas concentrates, 
and Fig.~\ref{fig-MMSN-rho} reveals a denser region of gas at the position of the planet which 
is harder to see in the vertically integrated surface density in 
Fig.~\ref{fig-MMSN-standard-2Ddens}, but which was hinted at by the reduced scale height of 
the gas disk in the lower left panel of Fig.~\ref{fig-MMSN-standard3D}.  The most striking feature 
of Fig.~\ref{fig-MMSN-rho} is that, because of the very efficient vertical settling, the dust 
volume density reaches that of the gas in several regions of the disk, 
and even exceeds it at the inner edge of the gap where the dust piles up. 
Dust may therefore significantly affect the structure and evolution of the disk in that regime.

In order to better understand the resulting structures in the dust disk, we ran a high
resolution 
($N$$\sim$400,000) simulation of our standard model and studied the 
particle loss rate. The simulations started with 200,000 gas particles and after 8 orbits the 200,000 
dust particles were added. The particle numbers were compared at the end of the simulation. (Note 
that some gas particles are lost in the first eight orbits and hence we only compare particle numbers 
from the eighth orbit onwards.)  
In Table~\ref{table-3} we tally the number of gas and dust particles lost 
when $R > R_{\rm esc}$, when $R=R_{\rm crit}$ (particles accreted by the planet), and when 
$R < R_{\rm inner}$ (particles accreted by the star).

\begin{table}
\caption{Particle loss rates from the 8th to the 104th orbit for high 
resolution runs of the standard MMSN model. Column one shows the accretion
zone where particles are lost and  columns two and three are the number of gas
and dust particles lost respectively (with the percentage in brackets). 
Row one shows the total number of particles (gas$+$dust) remaining at $P=104$.}
\label{table-3}
\centering
\begin{tabular}{lll}
\hline\hline
Accretion             & $N_{\rm gas}$  & $N_{\rm dust}$ \\ \hline
(Remaining)           &  \multicolumn{2}{c}{329,841 (87.5\%)} \\
{$R > R_{\rm esc}$ }  & 15,663 (4.2\%) &  202   (0.05\%) \\
{$R = R_{\rm crit}$}  & 18,746 (5\%)   &  3,180 (0.8\%) \\
{$R < R_{\rm inner}$} & 9,424  (2.5\%) &  116   (0.03\%) \\ \hline
\end{tabular}
\end{table}
%
\begin{figure}[h]
\resizebox{\hsize}{!}{
\includegraphics{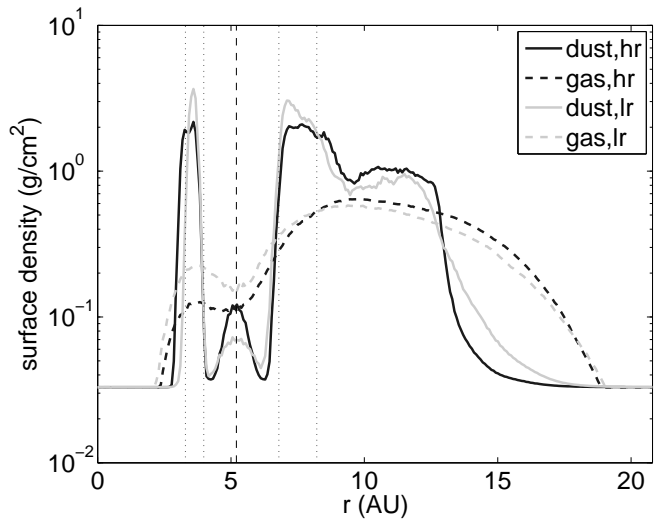}
}
\caption{The resulting gap for the standard model for the low resolution (lr: grey) and high resolution 
             (hr: black) simulations. Dust profiles are shown as solid lines while gas profiles (scaled by 0.01 
              to compare with the dust) are shown as dashed lines.}
\label{fig-MMSN-res}
\end{figure}

We find that about 12\% of the gas particles are lost while less than $1$\% of the dust 
particles are lost. This is because dust particles tend to move inwards at the outer disk edge
while gas particles tend to move outwards into the pressure void. At the edge of the planet 
gap, dust particles are trapped in the gas pressure maxima and so very few particles are actually captured 
by the planet.  At the inner disk edge, the gas particles again feel the pressure void and move 
inwards, while the dust particles remain where the gas pressure is highest.

Figure~\ref{fig-MMSN-res} compares the resulting gas and dust surface density profiles of the standard model for the low and high resolution simulations. The low resolution simulation 
captures the general shape and depth of the gap quite well, though it slightly overestimates 
the peak density at the gap edges and underestimates the dust resonant trapping by the planet.
We are confident that our low resolution simulations effectively capture the features of the planetary 
gap.

\subsection{Effect of grain size}
\label{subsec-grainsize}

As discussed in Sect.~\ref{sec-gapcriteria}, the scale height of the disk 
determines whether a planet of a given mass will open a gap in the disk,
and from the results of BF05 we know that the evolution of the dust disk's 
scale height depends on the grain size. 
BF05 found that large bodies ($s \ge 1$~m) are weakly coupled to the gas and 
essentially follow Keplerian orbits, while small bodies ($s \le 10\,\mu$m) are 
very strongly coupled to the gas. In both cases, the dust disk remains flared 
but for different reasons. Only in the inner part of the disk, where the gas 
density is highest and hence the gas drag is most efficient, does the disk 
flatten in the case of large grains. For intermediate sized grains 
(100~$\mu$m $\le s \le$ 10~cm) BF05 found that the drag strongly affects the 
dust dynamics, resulting in rapid settling to the midplane and subsequent 
strong inward radial migration. They found that 10~cm grains experienced 
the fastest radial migration.
While the nebula conditions of BF05 are not identical to those used in this study,
we expect the results to be qualitatively similar.


\begin{figure*}
\includegraphics[width=18 cm]{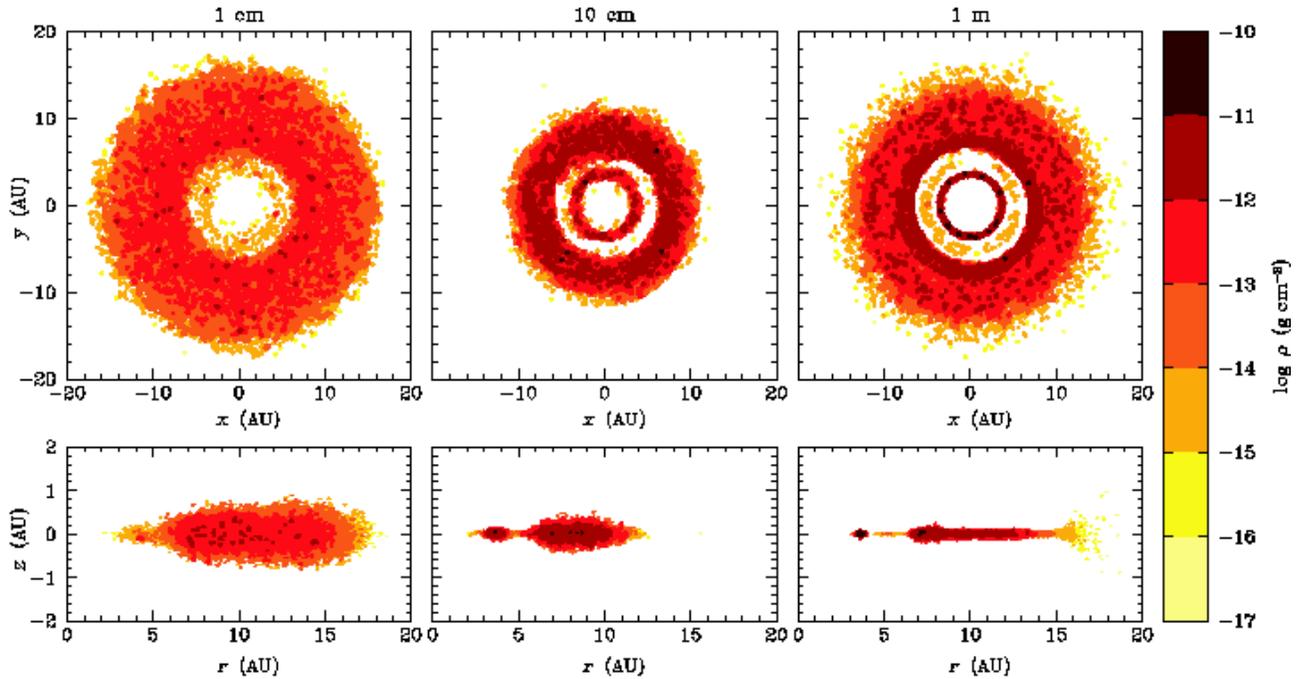}
\caption{Gap created in the dust disk for a range of grain sizes by a 
  1~$M_{\rm Jup}$ planet at 5.2~AU after 104 orbits.  Top panel shows the 
  top-down view of disk and bottom panel shows the side-on view. From left 
  to right: 1~cm, 10~cm and 1~m sized grains.  The dust is coloured by density.}
\label{fig-MMSN-grains3D}
\end{figure*}

\begin{figure}
\resizebox{\hsize}{!}{
\includegraphics{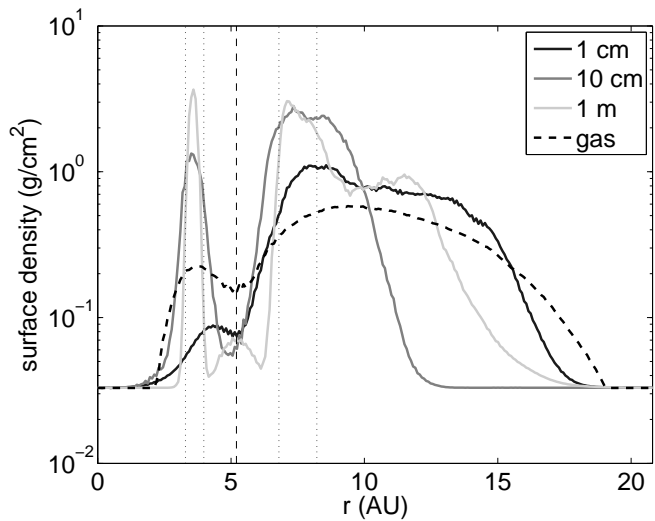}
}
\caption{Azimuthally averaged surface density profile for a gap created in the
  dust layer  for different grain sizes (1~cm, 10~cm, 1~m) by a 1~$M_{\rm Jup}$ 
  planet at 5.2~AU after 104 orbits.  Gas surface density is scaled 
  by 0.01 to compare with the dust and the vertical lines show the resonances 
  as in Fig.~\ref{fig-viscosity}. }
\label{fig-MMSN-grains-2Ddens}
\end{figure}

We ran a series of simulations to determine the 
effect of grain size on the induced planetary gap. Three grain sizes are 
tested: 1~cm, 10~cm and 1~m.
Figs.~\ref{fig-MMSN-grains3D} and \ref{fig-MMSN-grains-2Ddens} show how
the shape and depth of the gap in the dust phase of the MMSN disk varies according to the dust grains size.  (Note that the gas disk is unaffected by a change in grain size and looks identical to the left hand panel of Fig.~\ref{fig-MMSN-standard3D}.) 

\begin{figure}
\resizebox{\hsize}{!}{
\includegraphics{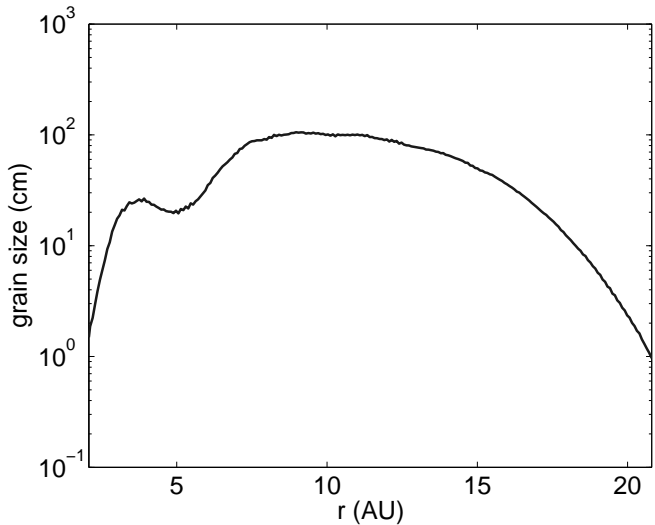}
}
\caption{Size of the grains with the largest radial migration velocity at a given radius in the disk.}
\label{fig-MMSN-stop}
\end{figure}

Similar to BF05, we find that the radial migration is most efficient for the 10 cm sized grains over most of the disk, which differs from the findings of Weidenschilling (1977). This discrepancy can be explained by the different nebula parameters used, in particular the density.
Indeed, \citet{Weidenschilling1977} noted that the radial migration velocity reaches a maximum for grains which have $t_{\rm s}/t_{\rm orb}\simeq 1/2\pi$, where 
$t_{\rm orb}=2\pi/\Omega_{\rm Kep}$ is the orbital period. Such grains have therefore $T_{\rm s}\simeq 1$ and, following Eq.~(6), one can estimate the ``optimal'' grain size for radial migration at a given location in the disk by
\begin{equation}
s_\mathrm{opt}=\frac{\rho_\mathrm{g}c}{\rho_\mathrm{d}\Omega_\mathrm{Kep}} .
\end{equation}
Figure~\ref{fig-MMSN-stop} displays the variation of $s_{\rm opt}$ with distance from the central star: decimetric grains are the most efficient radial migrators in the planet gap region, meter-sized grains migrate most efficiently from the outer gap edge out and decimetric grains take over again in the outer disk.  Centimetre grains are only efficient at migrating in narrow regions on the very edges of the disk.

As a result, while the 1~m boulders have the most rapid settling rate, the 10~cm grains have the most rapid radial migration rate in the outer disk and produce a small, truncated, high density dust ring exterior to the planet. The density increase for the 10~cm grains in the external disk is comparable to that in the case of 1~m boulders due to two competing effects: the settling is a little less efficient and leads to a thicker dust layer than in the 10~cm dust simulations, but more dust coming from the outer edge, where the radial migration is more efficient, accumulates at the external edge of the gap. The 1~cm grains are most strongly coupled to the gas and thus their timescale of settling is much slower. This results in a broader, lower density dust disk. While the inner region of the 1~cm-sized grain disk appears evacuated of dust (Fig.~\ref{fig-MMSN-grains3D}), this is due to the smaller density increase around the gap (Fig.~\ref{fig-MMSN-grains-2Ddens}) on the timescales shown. This is further supported by the lower accretion rate seen in the 1~cm grains than in the 10~cm grains.

As can be seen from Fig.~\ref{fig-MMSN-grains-2Ddens}, the gap gets deeper and 
wider as the grain size increases. Our findings are in contradiction to those of 
\citet{Paardekooper06} who claim that the grain size has no influence on the 
width of the gap (though they do note that the gap gets deeper with increasing 
grain size as we do).
While it is true that SPH will smooth the edges of the gap over $2h$, we do not believe that this alone is the cause of the grain size dependent gap width.  The $h$ values are about 0.1~AU around the gap edges in the 1~m and 10~cm cases and 0.2~AU in the noisier 1~cm case, which is far smaller than the smoothing of the gap edges seen in 
Fig.~\ref{fig-MMSN-grains-2Ddens}.
We assume, therefore, that this discrepancy is due to the evolution of 
the dust layer's scale height with grain size which we can follow in our 3D 
simulations. Within this size range, the dust disk scale height diminishes
as the grain size increases, and hence the planet opens a gap more rapidly
in larger grain sized dust layers. 

For the 1~m boulders case, particles are found in corotation with the planet.  
We would expect that some particles would be found at corotation for all grain sizes, but
we assume that this is only seen in the 1~m case because the gap is deeper and wider
than in the other two cases, making those particles caught in corotation easier to see.

\citet{Rice-etal2006} find that the pressure gradient at the outer gap edge 
act as a filter, allowing smaller particles to penetrate into the inner disk 
while holding back larger grains. Their transition size is $s\lesssim 10 \mu$m 
or smaller, and depends little on disk parameters. The dust grains we use in 
our simulations are all larger than this critical size and we indeed find 
that particles outside the gap do not migrate inwards, leaving the inner disk 
to drain by accretion on the central star.

\subsection{Effect of the planetary mass}
\label{subsec-planetmass}


\begin{figure*}
\centering
\includegraphics[width=18 cm]{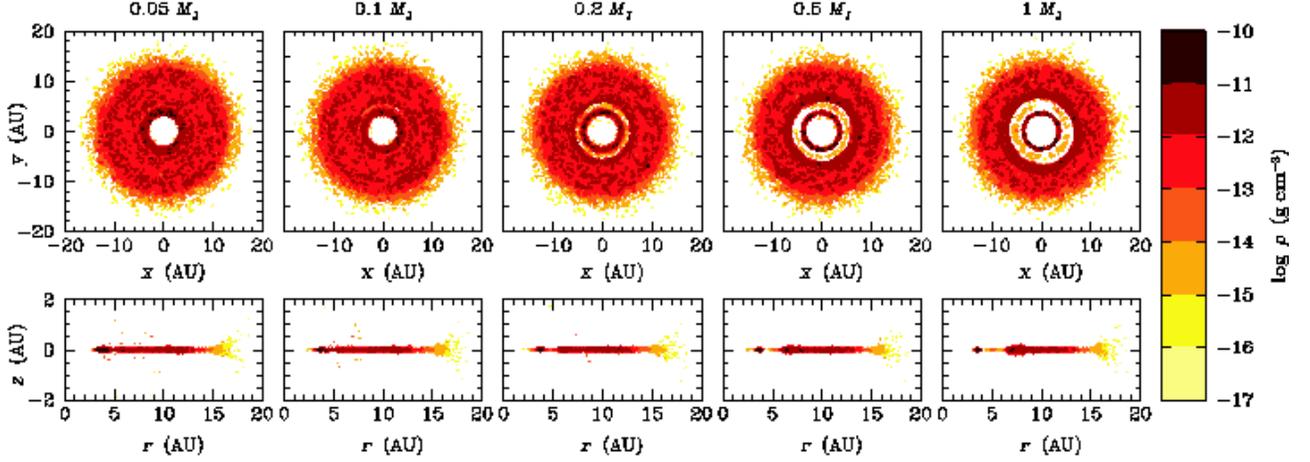}
\caption{Gap created in the dust disk comprising 1~m grains by a planet of varying 
  mass at 5.2~AU after 104 orbits. Top panel shows the top-down view of disk 
  and bottom panel shows the side-on view. From left to right: a planet of 
  mass 0.05, 0.1, 0.2, 0.5 and 1.0 $M_{\rm Jup}$. The dust is coloured by density.}
\label{fig-MMSN-planet3D}
\end{figure*}

We also investigate the effect of the planetary mass on the evolution of the
gap in the dust disk.  We return to our standard MMSN model which 
contains 1~m sized grains and vary $M_p$ from 0.05~$M_{\rm Jup}$ to 1~$M_{\rm Jup}$.
The resulting disk morphologies are shown in Fig.~\ref{fig-MMSN-planet3D}.
(Note that in this series of tests, the increasing planet mass does affect the 
gas disk distribution as well. However, the effect is quite small and so we do not include gas distribution figures.)

\begin{figure*}
{
\includegraphics[width=8.5cm]{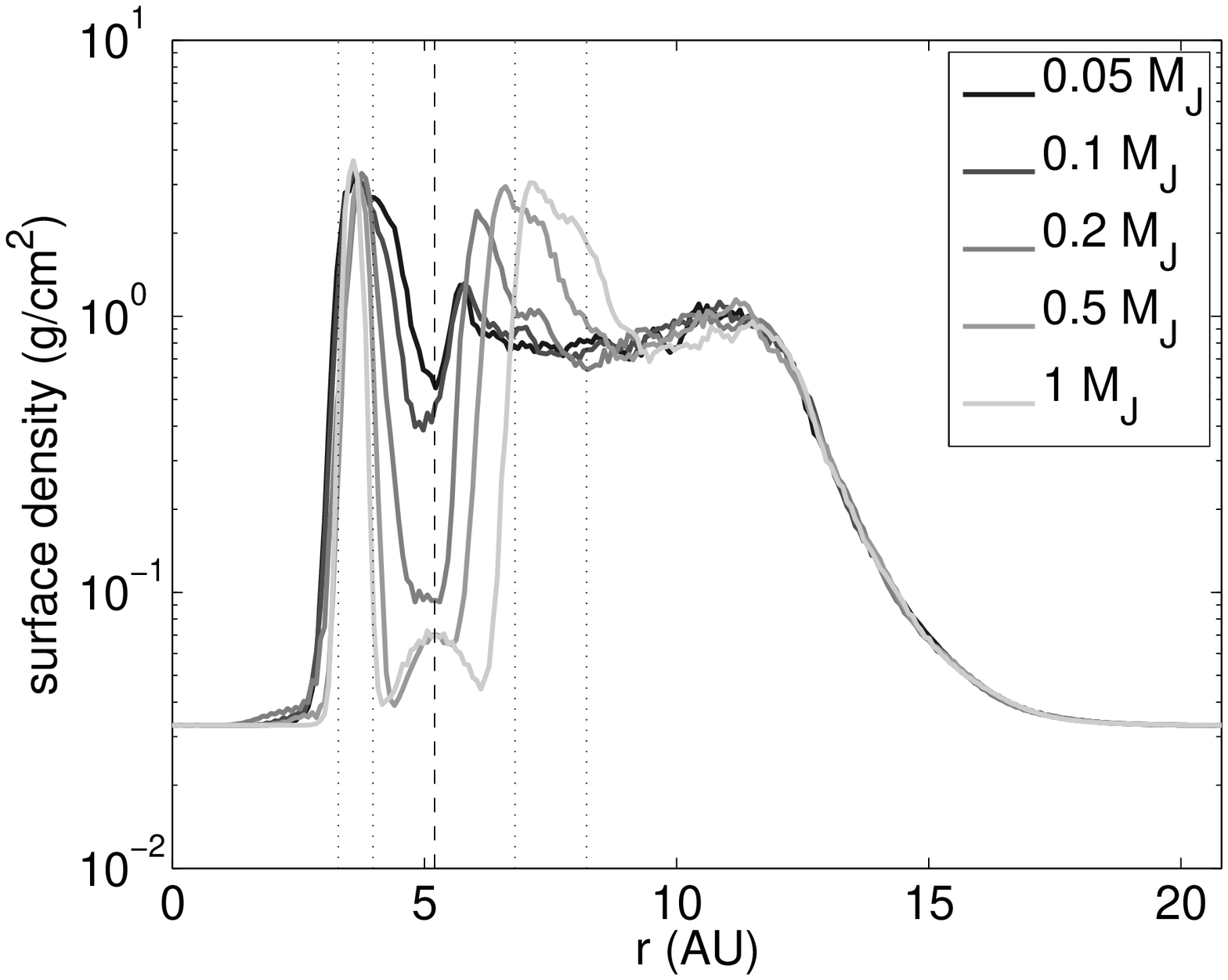}
\includegraphics[width=8.5cm]{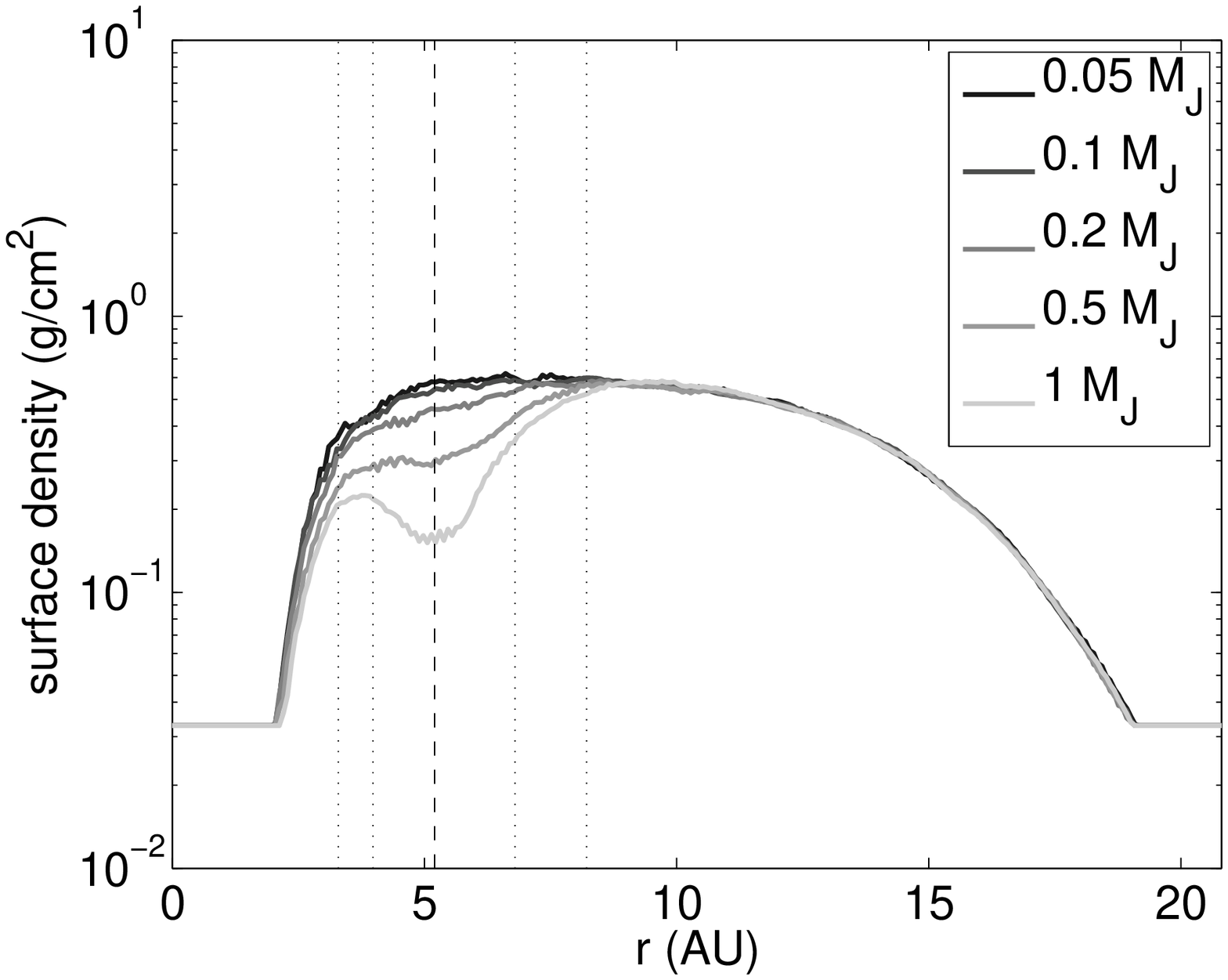}
}
\caption{Azimuthally averaged surface density profile for a gap created in 
  the disk after 104 orbits by planets of various masses 
  after 104 orbits.  The left panel shows the evolution of the dust 
  layer containing 1~m sized grains and the right panel shown the evolution
  of the gas layer which has been scaled by 0.01 for direct comparison with 
  the dust. The vertical lines show the resonances as in 
  Fig.~\ref{fig-viscosity}.  }
\label{fig-MMSN-planet-2Ddens}
\end{figure*}

The azimuthally averaged surface density profiles after 104 orbits are shown
in Fig. \ref{fig-MMSN-planet-2Ddens} for the dust and gas phases. 
An 0.05 $M_{\rm Jup}$ planet produces a decrease in the dust surface 
density in the vicinity of its orbit, whereas 
the gas surface density is not affected until the planet is more massive than 
0.2 $M_{\rm Jup}$.

As expected, the gap gets deeper and wider as the mass of the planet
increases. 
In the inner disk, the dust surface density increase is due to the dust accumulation 
at the gas pressure maximum.  While grains cannot cross the gap from the outer disk, in the inner disk grains move towards the pressure maximum coming from both sides of the maxima. The peak of the dust surface
density is dictated by the efficiency of this process and does not depend on the
gap depth.
However the density peak is broader for smaller mass planets
because of the correspondingly shallower gap. For low mass planets,
material can still flow from the outer to inner disk and therefore we have 
more mass in the inner disk than when the gap is deep enough to prevent 
this inflow. In the case of larger mass planets, the density peak gets 
thinner before its maximal value starts to decrease.

In the outer disk, the maximum surface density increases a little with the
mass of the planet but reaches almost the same value for the 0.2, 0.5 and 1
$M_{\rm Jup}$ planets. The peak is almost as broad in each case but is shifted 
to larger radii as the gap widens with increasing planetary mass.

For the two most massive planets modelled ($M_{\rm p}=0.5$ and 1 $M_{\rm Jup})$ 
some dust particles appear to become trapped near corotation. Once again these
are the simulations in which the gap is deepest and widest and hence it is easier 
to see particles in corotation.

In all cases but for the 1 $M_{\rm Jup}$ planet, the gap is again shifted 
slightly towards smaller radii than the planet orbit. 

\citet{Paardekooper06} find that the smallest planet mass able to open a gap in a 
dust disk comprising 1~mm-sized grains located at 13~AU is $0.05 M_{\rm Jup}$ for simulations 
of a MMSN disk. 
While their nebula parameters are not identical to ours and their models are 
2D and less viscous ($\alpha_{\rm SS}=0.004$), there is still general agreement between our 
results. From Eq.~(\ref{eqn-gapvisc}) we would expect a larger mass planet is 
required to open a gap at smaller radii, but with effective dust settling, the disk 
thickness is small at small radii which reduces the planet mass required to open a gap.
Thus we also find that an $0.05 M_{\rm Jup}$ planet at 5~AU (for our disk parameters) can 
open a gap in the dust disk.
%

\section{Discussion}

Structures created by planets in dusty disks are more diverse than those created in the gaseous disks.  With only aerodynamic drag, we already have the possibility of creating either a disk with a large central hole or a ring.  \citet{Rice-etal2006} also found that, for certain grain sizes, planets can create a central hole in the disk. Other processes can form similar structures, e.g. disk clearing by photoevaporation \citep{AlexanderArmitage2007}. The influence of radiation pressure and Poynting-Robertson drag would clearly lead to other more detailed structures for smaller grains 
\citep[see, e.g. ][]{TakeuchiArty2001}.

\subsection{The effect of the inner boundary}
\label{subsec-boundary}

Recently \citet{CridaMorbid2007} showed that the location of the inner boundary in planetary 
gap simulations can have a dramatic effect on the resulting structure of the innermost part of the disk.  
They found that a planet will open an inner cavity (i.e. an inner hole) in the disk if the planet is more 
massive that the disk or if the inner disk edge, $r_{\rm in}$, is too close to the planet.  They show that the 
disk surface density profile in the inner disk is lowered by a factor $1-\sqrt{r_{\rm in}/r_{\rm p}}$ in the 
presence of a planet.

\begin{figure}[h]
\resizebox{\hsize}{!}{
\includegraphics{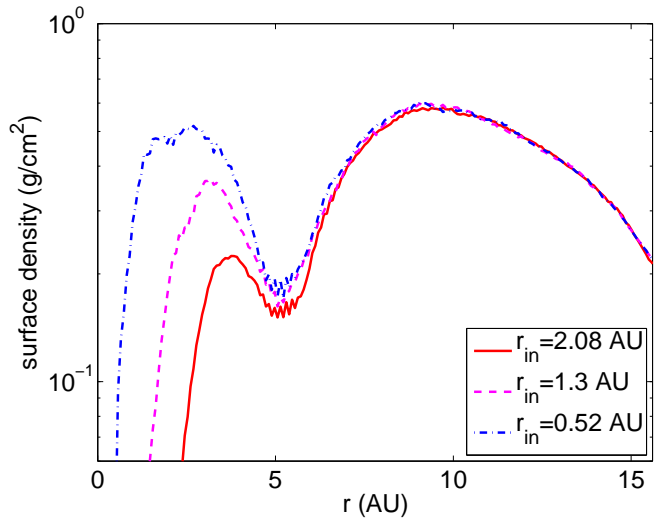}
}
\caption{Surface density profile of the gas in the standard model for three different inner boundaries: 
$r_{\rm in}=2.08$~AU, 1.3~AU and 0.52~AU.}
\label{fig-MMSN-innerBC}
\end{figure}
We re-ran our standard model with the inner boundary of the disk at $r_{\rm in}=0.1$ and 0.25 to compare the resulting 
inner disk structure with our $r_{\rm in}=0.4$ simulations.  Using the formula of \citet{CridaMorbid2007}, we
should find that the inner surface density drops by 32\% when $r_{\rm in}=0.1$, compared to a 63\% decrease when $r_{\rm in}=0.4$, which is what is seen in Fig.~\ref{fig-MMSN-innerBC}   
\citep[and compares well with Fig.~7 of][]{CridaMorbid2007}.

\citet{CridaMorbid2007} speculate about what would happen to dust in this inner disk region. They suggest that the dust would accumulate at the relative maximum of the gas density, which is indeed what we find, and that a thin dust ring would result, just as we see in our disks in Fig.~\ref{fig-MMSN-grains3D}. 

While it is clear that the location of the inner boundary has a dramatic effect on the density distribution of the inner disk, the results we have presented for different grain sizes and different planet masses should not change for a fixed
$r_{\rm in}$.

\subsection{Trapping in the 3$:$2 resonance}
\label{subsec-resonance}

\begin{figure}
  \resizebox{\hsize}{!}{
    \includegraphics[width=7cm,angle=-90]{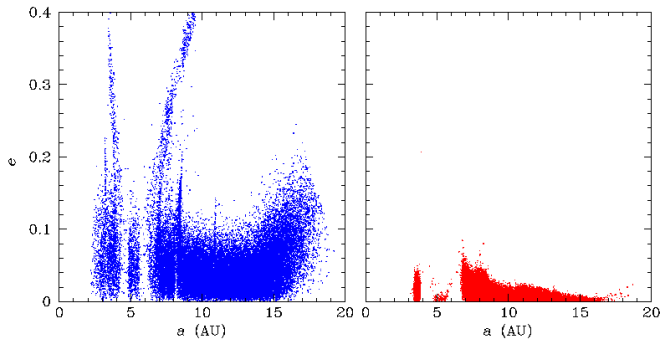}
  }
  \caption{Eccentricity ($e$) versus semi-major axis ($a$) for the standard MMSN disk with a 
  1~$M_{\rm Jup}$ planet orbiting at 5.2~AU after 104 orbits. Left panel: without gas drag. 
   Right panel: gas drag included.}
  \label{fig-ae}
\end{figure}

\citet{Paardekooper06} suggested that particles become trapped in the 3$:$2 
external resonance with the planet.  While we see a clear density increase 
in the vicinity of the 3$:$2 resonance of our standard MMSN disk in 
Fig.~\ref{fig-MMSN-standard-2Ddens},  it appears to be too broad to caused by resonant 
trapping.  To investigate further, we 
plot the eccentricity of the dust grains versus semi-major axis in 
Fig.~\ref{fig-ae}. 
In these kind of diagrams, resonances appear as thin vertical lines and as a 
V-shaped pattern at the edges of the gap. We ran a simulation without the drag 
force in which dust particles are fully decoupled from the gas. The results are 
shown on the left panel of Fig.~\ref{fig-ae} and the signatures of resonant 
trapping can be clearly seen. The rather high eccentricity of the dust particles 
is due to the initial conditions -- the dust was initially given the velocity 
field of the gas and without drag there is no mechanism to damp the gas 
velocity dispersion. In the right panel of Fig.~\ref{fig-ae}, we see that when 
simulations are run with gas drag, the drag is very efficient at damping the 
particles' eccentricity and consequently all resonant signatures disappear.
Furthermore, if we compare our results for different grain sizes (see Fig.~\ref{fig-MMSN-grains-2Ddens}) and for
different planet masses (Fig.~\ref{fig-MMSN-planet-2Ddens}), we see that the particle pile up at the 
outer edge of the gap does not always coincide with the 3:2 resonance for the same 
evolutionary phase (104 planet orbits).  This seems to depend quite sensitively on 
the exact disk parameters.

We conclude, therefore, that the accumulation of dust that we and other
authors 
\citep{Paardekooper06,AlexanderArmitage2007}
notice close to the outer gap edge is not due to resonant trapping.
Resonant trapping will depend on the efficiency of the gas drag for the grain sizes under consideration.

\subsection{Observations of protoplanetary disks}
\label{subsec-obs}

Our results have implications for recent predictions of observations of protoplanetary 
disks hosting planet gaps. As mentioned in Sect.~1, \citet{Wolf05} and 
\citet{Varniere-etal2006a} have used the results of 2D hydrodynamics simulations to produce 
synthetic images of protoplanetary disks 
following the suggestion of \citet{Paardekooper04} that such disks would be visible with ALMA. 
Such 2D simulations assume that the gas and 
dust are well mixed, which our results clearly demonstrate is not the case. 

The results presented here cannot directly be used to construct synthetic observations because we only
consider one discrete grain size per simulation. However, since we don't consider interactions between
dust grains and since the structure of the gas disk doesn't vary with grain size, we can stack our
results for different grain sizes to reconstruct a grain size distribution. With a finer size sampling
than presented here, we will be able to create synthetic scattered light images \citep[see][]{Pinte-etal2007}.

Because our general findings show than the gap is generally more striking in the dust 
disk, we suggest that predictions of observations of protoplanetary disks are too pessimistic. 
Our results show that the density contrast around the gap can be very strong  - up to four orders of
magnitude (see Fig.~\ref{fig-MMSN-rho}) - and this would be detectable. 
Our results, together with those of \citet{Paardekooper06}, show that even if one varies 
the nebula parameters, the result is robust: an $0.05 M_{\rm Jup}$ planet opens a gap in the 
dust disk.  As suggested by \citet{Varniere-etal2006a} the resulting strong density enhancements 
at the edges of the gap should be easily visible with ALMA.

While the interactions of the dust with the gas disk and the planet are complex, our 
simulations can help constrain the planet mass and grain size distribution in 
future high angular resolution observations of protoplanetary disks.

\subsection{Enhanced planetesimal growth}
\label{subsec-planetesimalgrowth}

We have shown that the presence of the planet triggers an
accumulation of grains at the external edge of the planetary gap, and the
resulting density enhancement may well favour the growth of planetesimals in
the region.  
This can be put in perspective with the results of \cite{Tsiganis05}, who 
propose a scenario for the formation of the Solar System that reproduces 
the characteristics of the giant planets as well as that of the Trojans 
and explain the late heavy bombardment of the Moon. This scenario relies 
on the controversial hypothesis that Jupiter and Saturn formed closer to one 
another than they are today and subsequently migrated, one inwards and the 
other outwards. While our results are rather extreme in that one of the 
planets is fully formed whereas the other is only at the planetesimal stage, 
even a planet embryo (well before it could carve out a gap) would create 
regions of pressure enhancement in its vicinity where dust would accumulate.
Such dense rings would be ideal sites for the concentration of solid
particles, and may lead to faster growth of planetesimals 
\citep{Durisen-etal2007}. 

\subsection{Planet migration}
\label{subsec-migration}

In this paper, we do not investigate the influence of planetary migration. 
The planet remains on a fixed circular orbit in our simulations and momentum  
exchange between planet and disk is not taken into account. 
We would expect, however, planet migration to have an effect on the dusty gap 
formation. Because the total dust mass is two orders of magnitude lower than 
the gas, one would expect the migration to be driven by gas and not dust. 
However, would this really be the case?  Ignoring the influence of turbulence, 
once the dust has settled to the midplane its density tends to be comparable 
to that of the gas (see Fig.~\ref{fig-MMSN-rho}). 
One may speculate that the 
planet might be locked to the dust evolution and gas drag implies that dust 
flows faster towards the central object than gas.  However, turbulence would 
need to be taken into account to fully study such a scenario. 
We will investigate the influence of migration in a dusty protoplanetary disk in a future paper.



\section{Conclusions}

We have conducted a series of 3D numerical simulations of two-phase 
(dust$+$gas) protoplanetary disks to study the behaviour of the dust in the 
presence of a planet. We ran a series of experiments with a minimum mass solar 
nebula disk, varying the grain size and the planet mass.

We find that gap formation is more rapid and striking in the dust layer than 
in the gas layer. Varying the grain size alone results in a variety of 
different structures which may well be 
detectable with ALMA. For low mass planets in our MMSN disk, a gap 
was found to open in the dust layer while not in the gas layer. 

Grains 
accumulate at the external edge of the gap but we find that this is not due 
to resonant trapping when gas drag is included. This is more likely due to 
the dust concentrating at the gas pressure maxima.  Contrary to previous 
2D work, our 3D simulations demonstrate that there is 
a dependence of the width and depth of the gap on grain size.
Simulations like these can be used to help interpret observations to 
constrain the planet mass and grain sizes in protoplanetary disks.

\begin{acknowledgements}
Computations presented in this paper were performed at the Centre Informatique National de
l'Enseignement Sup\'erieur (France). We thank the 
Programme National de Physique Stellaire of
CNRS/INSU, France, the Programme International de Cooperation Scientifique (PICS) France-Australia in 
Astrophysics (Formation and Evolution of Structures), and the Swinburne University Research 
Development Grant Scheme for partially supporting this research. 
We also thank the anonymous referee whose comments have greatly improved this paper.
\end{acknowledgements}

\bibliographystyle{aa}
\bibliography{Fouchet-etal}

\end{document}